\documentclass[apj]{emulateapj}

\usepackage{color}

\usepackage{natbib,graphicx}

\usepackage{epsfig}
\usepackage{ulem}
\usepackage{color}
\usepackage{graphics,graphicx}
\usepackage{times} 
\usepackage{amssymb}
\usepackage{amsmath}
\usepackage{url}
\usepackage{multirow}
\usepackage{ctable}
\usepackage{threeparttable}
\newif\ifAMStwofonts
\AMStwofontstrue

\def\apj{ApJ}
\def\mnras{MNRAS}
\def\nat{Nat}

\def\araa{ARA\&A}                
\def\aap{A\&A}                   
\def\aj{AJ}                      
\def\apjl{ApJ}                   

\def\araa{ARA\&A}                

\def\epicmos1{{\it EPIC}{\rm-MOS1~\/}}
\def\epicmos2{{\it EPIC}{\rm-MOS2 ~\/}}
\def\epicmos{{\it EPIC}{\rm-MOS}}

\def\msun{\hbox{$\rm\thinspace M_{\odot}$}}

\def\rg{${\it r}_{\rm g}$}

\def\grid25{\hbox{\rm{\small GRID25}}}

\def\pile_est{\hbox{\rm{\small PILE_EST}}}

\def\j1118{\hbox{\rm XTE J1118+480}}
\def\j1749{\hbox{\rm J17497-2821}}

\def\jb{\hbox{\rm XTE~J1650--500}}

\def\j1752{\hbox{\rm XTE~J1752--223}}

\begin{document}
\title[] {On the size and location of the  X-ray emitting coronae around black holes} 
\author{
R.~C.~Reis\altaffilmark{1,2},
J.~M.~Miller\altaffilmark{1},
}
\altaffiltext{1}{Dept. of Astronomy, University of Michigan, Ann Arbor, Michigan~48109, USA}
\altaffiltext{2}{Einstein Fellow}

\begin{abstract}
The observation of  energetic X-ray emission from black holes,
inconsistent with thermal emission from an accretion disk, has long
indicated the presence of a ``corona" around these objects.  However,
our knowledge of the geometry, composition, and processes within black
hole coronae is severely lacking.  Basic questions regarding their
size and location are still a topic of debate. In this letter, we
show that for  black holes with luminosities $L\gtrsim10^{-2}L_{Edd}$
-- characteristic of many Seyferts, quasars, and stellar-mass black
holes (in their brighter states) -- advanced imaging and timing data
strongly favor X-ray emitting regions that are highly compact, and only a
few Gravitational radii above the accretion disk. The inclusion of a large number of possible systematics uncertainties does not significantly change this conclusion with our results still suggesting emission from within $\sim20$\rg\ in all cases.   This result favors
coronal models wherein most of the hard X-ray emission derives from
magnetic reconnection in the innermost disk and/or from processes in the
compact base of a central, relativistic jet.

\end{abstract}

\begin{keywords} {}\end{keywords}

\section{Introduction}

Observations of accreting black holes (BHs) often indicate the presence of hard X-rays having energies much greater than the expected thermal peak of the accretion disk. This is true not only for Active Galactic Nuclei \citep[AGNs;~e.g.][]{Elvis1978} but also for X-ray binaries \citep[XRBs;~e.g.][]{White1982coronae}.

It has been widely postulated that  hard X-rays  are the product of inverse Compton scattering of seed photons from accretion disks by  hot ``coronae"  \citep[e.g.][]{HaardtMaraschi91}, however despite our observational \citep[e.g.][]{Zdziarski1999} and theoretical  \citep[e.g.][and references therein]{Schnittman2012states} efforts,  there are still a number of fundamental questions regarding the geometry, or  composition of  accretion-disk coronae.

At  low luminosities ($L\lesssim 10^{-4}L_{Edd}$, where  $L_{Edd}$ is the Eddington luminosity), the hard-X-rays are thought to originate from a hot, quasi-spherical and large  ($\sim100 GM/c^2 \sim 100$\rg) advection-dominated accretion flow (ADAF) which replaces the  innermost  disk around the BH \citep[e.g.][]{NarayanYi1994}. On the other hand,  at luminosities $L\gtrsim 10^{-2}L_{Edd}$, observations of XRBs and Seyfert AGNs show the clear presence of a more compact X-ray emitting source (the corona; e.g.~\citealt{McHardy2006}) on top of a geometrically-thin, optically thick accretion disk  extending as far in as the radius of the  innermost stable circular orbit (ISCO).  

Amongst the clearest evidence for the co-existence of the corona with the  disk is the presence of Fe K-shell emission lines together with other reflection features in the X-ray spectra of these sources \citep[][and references therein]{miller07review}. From the profile of these emission lines we are  able to obtain information on the effects of strong Doppler and gravitational redshifts natural to the space-time around BHs and measure their spin \citep[see][for a review]{miller07review}.

Having a tool to probe the innermost regions around BHs and to measure  their spin  ($a=cJ/GM^2$, $-1< a < 1$)  is of particular importance as the distribution of spin in nearby supermassive black holes (SMBHs) can be used to distinguish between various  BH  growth scenarios \citep[e.g.][]{BertiVolonteri2008}, and in the case of stellar mass BHs,  their spin distribution relates to the distribution of angular momentum of the progenitors   (e.g. \citealt{millersn11}). Knowledge of spin  is also essential if one wishes to test if relativistic jets can be powered by the spin energy of BHs via the Blandford-Znajek process \citep[e.g.][]{NarayanMcClintock2012} or, in the near future, if  there is evidence for violations of the No-Hair Theorem \citep[e.g.][]{Johannsen2011nohair}.

The location and compactness of the corona --  which is irradiating the disk and  giving rise to  the reflection features --  are  thus of  paramount importance in the study of strong gravity. Nonetheless,  in theoretical consideration of the corona it is often \textit{assumed} to be compact and located in the inner regions, $\lesssim10$\rg\ around the BH \citep[e.g.][and references therein]{Miniu04}. This assumption has consistently been successful in describing the spectral and timing properties of BHs at the highest signal-to-noise  available, however strong deviation from this could affect our ability to probe strong gravity around BHs \citep{Dauser2013}.

Furthermore, coronae are known to play an important role in state transitions in  XRBs \citep[e.g.][and references therein]{reis20121650}; and jets in both XRBs and AGNs have often been strongly linked with the corona \citep[e.g.][]{fundamentalplane}.  If  the corona is  the base of such a jet \citep{FalckeBiermann1999}, then the amount and the form in which energy escapes through the jet is largely dependent on the enigmatic properties of the corona, highlighting the importance in establishing the size and location of these regions.

In this letter, we explore the results aimed at measuring the physical sizes and locations of coronae around luminous accreting SMBHs ($L\gtrsim 10^{-2}L_{Edd}$ typical of  bright nearby Seyferts and quasars). We find that in all cases the coronae are located significantly closer than $\sim10$\rg\ from the accretion disks and their sizes are suggestive of being highly compact. These results give credence to the assumptions often employed in most reflection-based studies of BHBs and AGNs.

\section{Location based on reverberation lags}

With the discovery of a soft X-ray reverberation lag of $\sim30$s in
the AGN 1H~0707-495 \citep{FabZog09}, a new method to measure the
distance ($D_{\rm cor}$) between the X-ray emitting region and the accretion disks   was
established.  The decomposition of the spectrum of 1H~0707$-$495 into
direct and reflected emission components supported interpreting the
observed lag as a signature of the light travel time between the
corona (the hard X-ray continuum source) and the accretion disk (which
reprocess the flux into a reflection component with atomic spectral
features). 

Despite the success that this ``reflection" model had in explaining the timing and spectral features of 1H 0707-495 \citep{FabZog09}, a further interpretation was soon made available where the lag was due to the reverberation caused by scattering of X-rays passing through an absorbing medium that happen to lie in the line of sight to the central AGN \citep{MillerTurner2010lag}.  However, the subsequent discovery of similar lags in a large number of sources (see below), spanning a large range of masses, Eddington fractions, and inclination angles, is likely incompatible with the specialized partial covering geometry required to explain the lags. Thus, the original light-crossing argument remains the best interpretation for these soft lags.   For a detailed description of soft X-ray reverberation techniques and its applications  see \citet[][]{Zoghbi2012lag,WilkinsFabian2013}.

At the time of writing, there were a total of 17 radio-quiet Seyferts\footnote{The LLAGN NGC~4395 is not used in this work as it accretes constantly at $L/L_{Edd}\sim1.2\times10^{-3}$ and is possibly already in the ADAF  regime and thus  intrinsically different from the more luminous systems.} with measured soft lags that met our luminosity criteria. This sample is shown in Table~1 together with estimates for the  mass and luminosity as a fraction of Eddington to each source collected from the literature. The lags compiled here are taken from the work of \citet{DeMarcolags, Zoghbi2012lag, Fabian2012iras13224, Cackett2012ESOlag} and \citet{Kara2013iras},  whereas the masses are primarily  from \citet{zhouzhangmasses2010, Zhoumasses2005, Bentzngc41512006} and \citet{Onkenngc41512007}.  As Seyferts are known to be highly variable, the luminosities shown here are meant to be seen only as a rough indication of the average luminosity of each system.

\begin{figure}[!t]
\vspace*{-0cm}
\centering
{\hspace{-0.0cm} \rotatebox{0}{{\includegraphics[width=8.7cm]{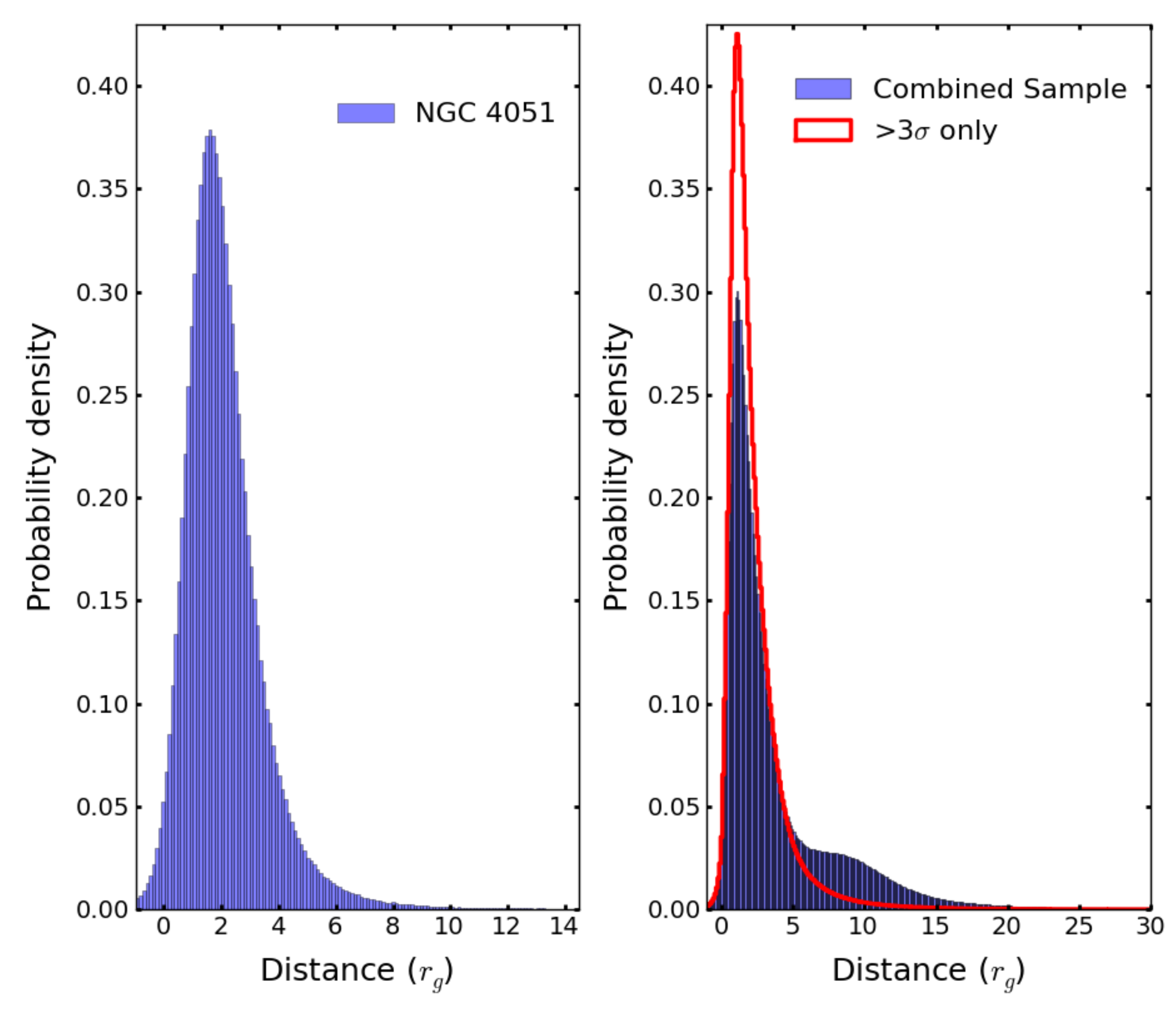} }}}
\vspace*{-0.cm}
\caption{ Left: Representative probability distribution for the distance between the accretion disk and the corona  ($D_{\rm cor}$) based on soft X-ray reverberation lags.  Right: Probability distribution for the combined sample of 17 Seyferts (blue). The tail extending to   $\sim15$\rg\ comprises of RE~J1034+396, NGC~4151 and ESO~113-G010.   Shown in red is the distribution for  sources with lag detection at $>3\sigma$, with the exception of NGC~4151. }  
\label{fig1}
\end{figure}

Due to the scale-invariant nature of the  physics governing accretion flows onto BHs, the lag time scale $t$ depend primarily on the mass of the central BH\footnote{This simple formula does not take into account the effect of the Shapiro time delay; however, this correction is of the order of 0.5\rg.  The apparent distance between the corona and reflector, calculated using the simplified formula and neglecting the Shapiro delay, is therefore larger than the actual distance (\citealt{WilkinsFabian2013} for details).}: $t\sim GM/c^3 = r_g/c$. We estimate the probability distribution  for the distances between the coronae and accretion disks by drawing samples from the probability distribution of the lag times and masses as obtained from the literature and  reported in Table~1.  For NGC~4051, MRK~766, RE`J1034+396, NGC~7469, Mrk~335, PG~1211+143, NGC~3516, NGC~5548 and  NGC~4151, their masses were assumed to have a Gaussian distribution with standard deviation given by the average of the positive and negative errors shown in Table~1, when the reported errors were asymmetric. For the remaining sources with the exception of ESO-113-G010, the logarithm of the masses were assumed to have a Gaussian distribution  with the symmetric errors reported in each reference. Finally, for ESO-113-G010, we assumed a uniform distribution in the range given in Table~1.

\begin{figure}[!t]
\hspace*{-0.3cm}
\centering
{\hspace{-0.0cm} \rotatebox{0}{{\includegraphics[width=8.7cm]{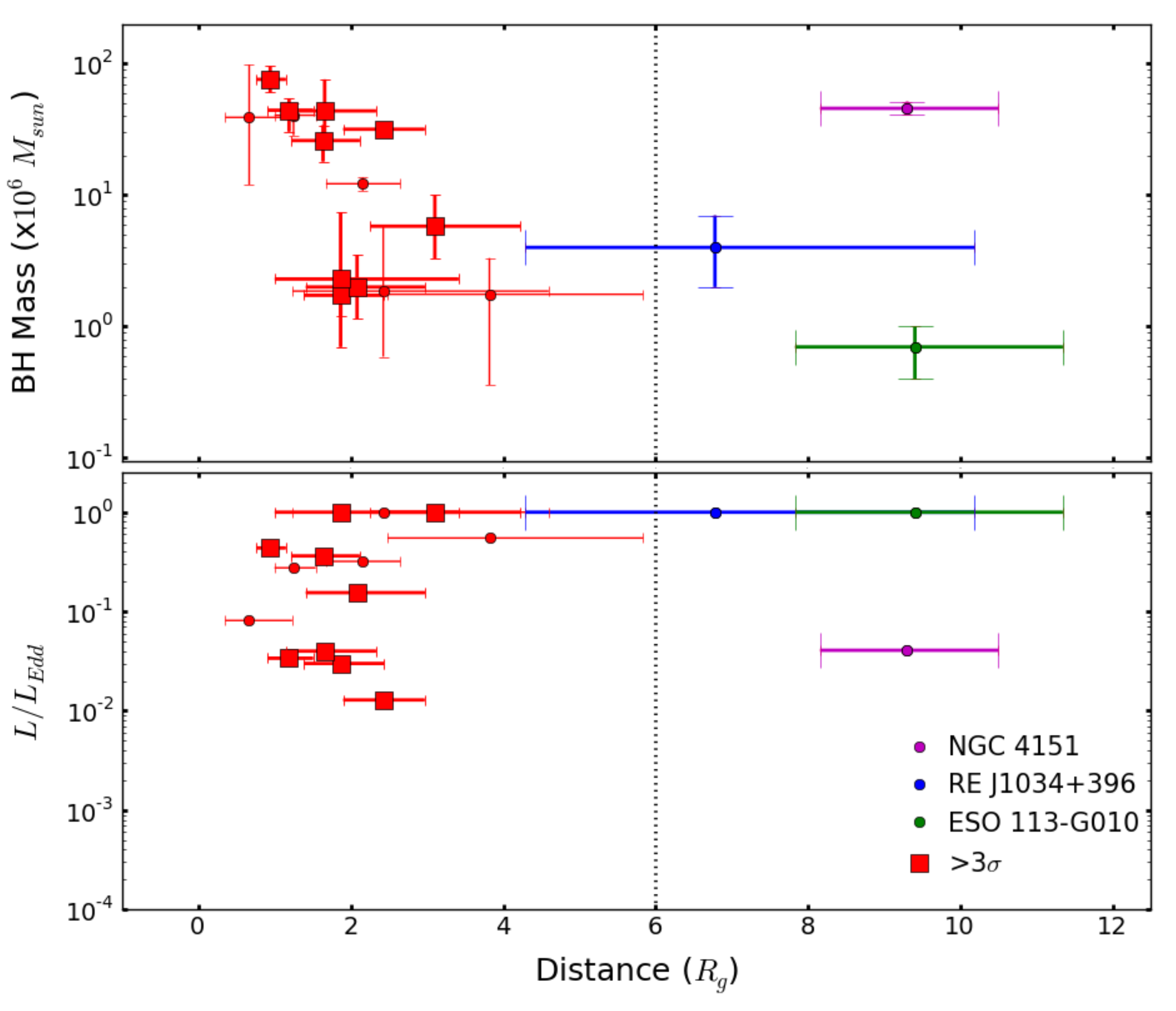} }}}
\vspace*{-0.1cm}
\caption{BH mass (top) and $L/L_{Edd}$ (bottom) as a function of disk-corona distance. The dashed line marks the ISCO of a non-rotating BH (6\rg). 14/17 sources in our sample suggest a corona  close to the accretion disk. These results are independent of the mass or $\dot{m}$ as long as the system are accreting above a few percent of Eddington. The three singled out sources corresponds to those peaking at  $\sim9$\rg\ (\S~2) and the  sources with lag detection at $>3\sigma$ level  are shown with  square markers.}  
\label{fig2}
\end{figure}

 \begin{table*}[!t]
 \parbox{\textwidth}{
\small 
\textbf{Table~1}: {Summary of sources used in this work. The table is divided into two with the top half referring  to the 17 Seyferts and the bottom half to  the 5 microlensed quasars. The various columns are (1) Object name; (2) black hole masses in Solar units. For the quasars, their masses are in logarithmic scale as quoted in the literature; (3) representative Eddington fraction;   (4) lag time in seconds (top) or logarithmic of the size of the X-ray emitting region in centimetres (bottom); (5) The median of the distance, $<D_{\rm cor}>$, between the corona and the accretion disk in gravitational units (top) or the median of the size of the corona $<S>$ also in gravitational units (bottom).   }}
 \label{table1}
   \begin{center}
   \begin{small}
    \begin{tabular}{lccccc}
\hline 
\hline
Object  & Mass ($\times 10^6\msun$) &$L/L_{Edd}$ &  Soft Lag Time (s) & $<D_{\rm cor}>$ (\rg)  \\ 
(1)						&(2)											& (3)							&(4)					& (5)			  \\
 \hline

NGC~4051 & $1.73^{+0.55}_{-0.52}$& 0.03 & $16\pm8$ & $1.9^{+0.6}_{-0.5}$   \\ 
 
Mrk~766 & $1.76^{+1.56}_{-1.40}$ & 0.55&$43\pm18$ & $3.8^{+2.0}_{-1.3}$  \\ 
 
Ark~564 & $1.86^{+4.03}_{-1.27}$ &1 & $26\pm15$ & $2.4 ^{+2.2}_{-1.2}$ \\ 
 
MCG-6-30-15 & $2.0^{+1.5}_{-0.9}$&  0.15&  $22\pm11$ & $2.1^{+0.9}_{-0.7}$  \\ 
 
1H~0707-495 & $2.3^{+5.1}_{-1.6}$&1 & $24\pm11$ &$ 1.9^{+1.5}_{-0.9}$   \\ 
 
RE~J1034+396 & $4^{+3}_{-2}$&1& $150\pm90$ &$ 6.8^{+3.4}_{-2.5}$   \\ 
 
NGC~7469 & $12.2\pm1.4$ & 0.32&$130\pm60$ & $2.2\pm0.5$  \\ 
 
Mrk~335 & $26\pm8$ &0.36 & $210\pm100$ & $1.6^{+0.5}_{-0.4}$   \\ 
 
PG~1211+143 & $40.5^{+9.6}_{-12.1}$ & 0.28 & $250\pm90$ & $1.3^{+0.3}_{-0.2}$  \\ 
 
NGC~3516 & $31.7^{+2.8}_{-4.2}$ &0.013& $380\pm170$ & $2.4\pm0.5$   \\ 
 
Mrk~841 & $75.9^{+19.6}_{-15.6}$ &0.44& $360\pm130$ & $0.9\pm0.2$  \\ 
 
NGC~5548 & $44.2^{+9.9}_{-13.8}$ &0.034& $260\pm115$ & $1.2\pm0.3$   \\ 
 
NCC~6860 & $39^{+59}_{-27}$ &0.08 & $145\pm72$ & $0.7^{+0.6}_{-0.3}$   \\ 
 
Mrk~1040 & $43.7^{+32.2}_{-18.5}$ & 0.04& $380\pm170$ & $1.7^{+0.7}_{-0.5}$   \\ 
 
NGC~4151 & $ 45.7^{+5.7}_{-4.7}$ &0.04& $2100\pm500$ & $9.3\pm1.2$  \\ 
 
IRAS~13224-3809 & $5.8^{+4.2}_{-2.5}$& 1& $92\pm31$ &$ 3.1^{+1.1}_{-0.8}$   \\ 
 
ESO~113-G010 & 4--10 & 1& $325\pm89$ & $9.4^{+1.9}_{-1.6}$  \\ 
 \hline
 \hline
 \\
 Object  &  log($M_{BH}/\msun$) & $L/L_{Edd}$& Log(Size/cm) & $<S>$ (\rg)  \\ 
 (1)						&(2)											& (3)							&(4)			&(5)					\\
\hline
Q~2237+0305 & $8.68\pm0.36$ & 0.44 &$15.46^{+0.34}_{-0.29}$ & $41^{+28}_{-17}$    \\ 
 
RX~J1131-1231 & $ 8.32\pm0.62$ & 0.03 & 14.04--14.68 & $7.4^{+7.6}_{-3.8}$  \\ 
 
Q~J0158-4325 & $8.2\pm0.2$ & 0.4 & $14.3^{+0.4}_{-0.5}$ & $8.5^{+6.1}_{-3.5}$    \\ 
 
HE~1104-1805 & $9.37\pm0.33$ & $0.36$ & 14.2-15.0 & $1.1^{+0.7}_{-0.4}$   \\ 
 
HE~0435-1223 & $8.76\pm0.44$&0.11 & $<15.07$ & $13.8^{+8.6}_{-5.3}$  \\ 

PG~1115+080 & $9.1\pm0.2$& $0.37$& $15.6^{+0.6}_{-0.9}$ & $21^{+29}_{-12}$   \\
\hline 
  \hline 
\end{tabular}
   \end{small}
 \end{center}
\parbox{\textwidth}{
\begin{footnotesize}
%

\end{footnotesize}

}  
\end{table*}

An example of a derived  probability distribution based on $10^6$ samples is shown in Fig.~1 (left) for NGC~4051. We report on the 5th column of Table~1 the median of  $D_{\rm cor}$ with the $1\sigma$ error taken to be the standard deviation of each source sample. Figure~1 (right) shows the distribution for the combined sample, where it becomes clear that there is a peaks at $\sim2$\rg\ and a further, minor peak at $\sim9$\rg\ with a tail that extends up to $\sim15$\rg. 

The 3 sources giving rise to the second peak at larger $D_{\rm cor}$ are NGC~4151, RE~J1034+396 and ESO~113-G010. There is currently uncertainty on the  mass of the  SMBH in NGC~4151. The  mass  determined from reverberation mapping is 4.57$^{+0.57}_{-0.47}\times 10^7~\msun$ \citep{Bentzngc41512006}. A stellar dynamical mass measurement obtained using ground based I-band long-slit spectra and the Schwarzschild orbit superposition code of  \citet{vallurietal04} gave only an upper limit of $\sim 4\times 10^7~\msun$ \citep{Onkenngc41512007}.  However, a recent result based on high resolution near infrared integral field spectroscopic observations of the CO-band head with the Gemini North NIFS spectrograph, gives a revised stellar dynamical mass of $(8.5\pm 3)\times 10^7\msun$ (M. Valluri; private communication) -- nearly a factor of 2 greater than previous estimates. If this turns out to be right,  $D_{\rm cor}$ for this source presented in Fig~2 should be closer to 5\rg.

As for the other two sources peaking at $\sim9$\rg, it is possible that they were indeed caught at a time when their coronae were  further than the median. It is interesting that RE J1034+396 is the only persistent AGN to ever show a quasi-periodic oscillation \citep[QPO;][]{rej1034qpo}. In the stellar-mass BH \jb, it has recently been shown that the onset of the more coherent QPOs are associated with the collapse of a corona from $\sim10$\rg\  during the transition from the hard-intermediate to the soft-intermediate state \citep{reis20121650}.\footnote{The only other QPO detected  from a SMBH comes from a tidal disruption source \citep{Reis2012qpo} and is  not persistently active.}

\section{Size based on microlensing}

The possibility of using the ``lens-like" effect of stars on distant object is not new.  However, its practical use in directly probing the inner regions around SMBHs  was not established until the discovery of a flux-ratio anomaly in  optical macroimages of the gravitationally lensed quasar QSO~2237+0305 where the optical flux ratio between the brightest macroimages was statistically  inconsistent with unity  \citep{Irwin1989}{\footnote{Under a smooth gravitational potential, without the effect of micro-lensing, the brightest image-pairs are expected to be mirror-images of each other and  have similar intensities \citep[][]{MetcalZhaof2002}.}}. Although originally microlensing was not the only explanation for the flux anomaly problem in the optical \citep[e.g.][]{DalalKochanek2002}, the systematic detection of an increase in the anomaly magnitude at higher energies strongly favoured microlensing as the cause of these anomalies \citep[e.g.][]{PooleyBlackburne2007}. As the effect of microlensing increases as the size scale of the lensed region decreases -- as long as the source is smaller than the projection of the Einstein radius of the  microlens into the plane  (for a detailed review of microlensing techniques  see \citealt{Wambsganss2006}) -- this discovery  confirmed the expected hierarchy in the characteristic sizes of the optical, UV and X-ray emission regions, with the latter often thought to be the most compact.

Microlensing studies at optical/IR have often arrived at accretion disk sizes in the order of $10^4$\rg\ \citep[e.g.][]{BlackburnePooley2006}, and most remarkably, such studies have shown that this radius scales with BH mass in a manner consistent with the expectations of thin-disk theory \citep{MorganKochanek2010}. We are now at a stage where the size of the  X-ray emitting corona in the innermost regions around these BHs  are beginning to be measured \citep[e.g.][and references therein]{sizepg1115, size1104, Mosquera2013}. We list in Table~1 (Bottom)  the details of the  6 sources with a quantitatively derived  half-light radius for the X-ray emitting region. These values were compiled from the various  work listed above and especially from Fig.~8 of \citet{Mosquera2013}.  As these are all SMBHs, the accretion disk emission peaks in optical/UV, so the size of the  X-ray emitting region therefore directly relates  with the  corona.\footnote{The use of half-light radius in microlensing studies is of particular importance as these have been shown to be independent of the  shape or the surface brightness profile of the source (e.g. \citealt{Mortonson2005} for details).} In compiling the various masses presented in Table~2, we have searched the literature for the latest  BH mass estimates based on the Mg$\rm {II} (\lambda2798\AA)$ mass-line width relation for all sources but HE~1104-1805, where C$\rm IV (\lambda1549\AA)$ was used instead. The values quoted are from the work of \citet{Peng2006amass} and  \citet{Sluse2012mass}.

In a similar manner to \S~2, we have estimated  a  probability distribution for the size of the X-ray emitting region of each source in units of \rg\ based on  $10^6$ randomly sampled values of mass and size (in cm) assuming  either a Gaussian distribution with symmetric errors in logarithmic scale or, in the cases where such are not present, a uniform distribution over the range given in Table~1. Figure~3 summarises the result for the characteristic sizes of the coronae.

\begin{figure}[!t]
\hspace*{-0.3cm}
\centering
{\hspace{-0.0cm} \rotatebox{0}{{\includegraphics[width=8.5cm]{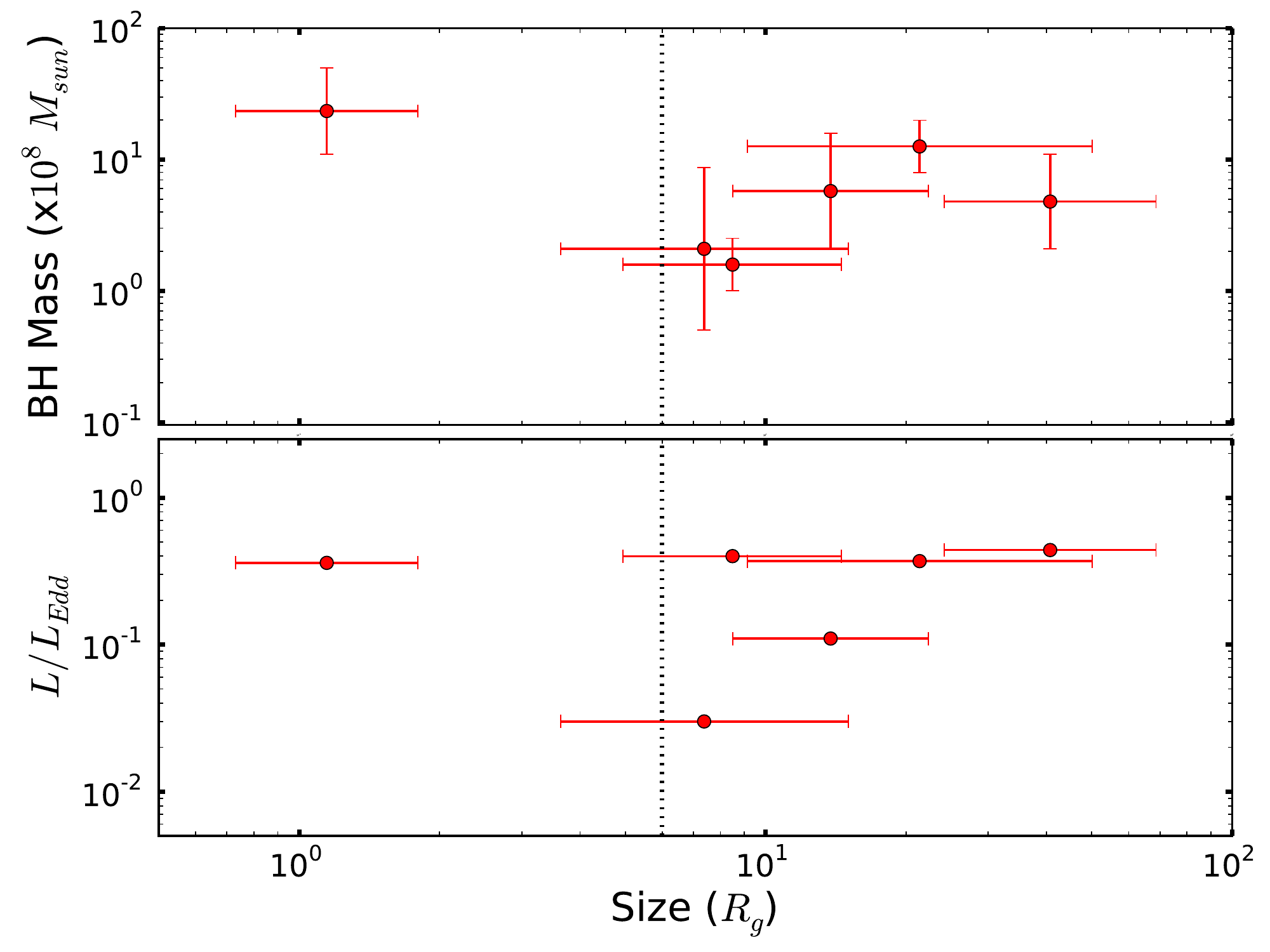} }}}
\vspace*{-0.3cm}
\caption{ Characteristic size of the X-ray emitting region  versus mass (top) and characteristic luminosity (bottom).   }  
\label{fig3}
\end{figure}

\section{Discussion}
 
The two techniques explored in this work effectively probe different
aspects of the same emission region. From the reverberation lags, we
show in Fig~2 that 14/17 sources in our sample\footnote{15/17 using
  the higher mass estimate of NGC 4151.}  are at distances strongly
clustered around 2--3\rg\ above the accretion disk. Taken at face
value and assuming that the corona is radially symmetric about the
central BH, this immediately suggest that a large fraction of Seyferts
should be rapidly spinning (as a point of reference, a radius $R_{\rm ISCO}=3.15$\rg\ corresponds to a spin parameter 
$a=0.75cJ/GM^2$), as is often found with detailed spectral analyses
\citep{Waltonreisspin2013}.  For a few sources, notably Mrk~841 and
NGC~6860, the coronae appear to be at distances commensurate with
radii for which there are no stable orbits, even for a maximally
rotating BH\footnote{Again under the assumption that the corona is
  radially symmetric about the central BH.}. However, the effect of
lag dilution could artificially reduce the intrinsic lag
\citep[e.g.][]{WilkinsFabian2013} thus
increasing the distance between the disk and
corona{\footnote{\citet{WilkinsFabian2013} showed that this effect
    could increase $D_{\rm cor}$ in 1H~0707-495 by up to 75\%, and a factor of similar magnitude  could also affect all other lags mentioned here.}}; and
possible errors in the BH mass could also change the absolute $D_{\rm cor}$
reported here.

 A number of other systematics uncertainties in the derivation of soft-lags, including the unknown geometry of the corona or intrinsic flux variation \cite[e.g.][]{Zoghbi2012lag,Kara2013iras} could also increase the  radius by a factor of a few. Indeed, for the case of NGC~4151, \citet{Zoghbi2012lag} showed that the lag-time differed by a factor of $\sim10$ between high and low flux states. The result presented here conservatively  shows the upper value ($\sim 2000$~s), as found in the low-state of the source. If instead, we use the $\sim200$~s lag, the distance between the disk and the corona would  commensurate with a single gravitational radii. It is clear that for the majority of the Seyferts reported here, the distance between the X-ray emitting region and the inner 
accretion disk is  smaller than $\sim20$\rg\ even if we allow for an order of magnitude systematic shift in the derived values.  The microlensing
results further suggest that these coronae are highly compact, with
upper limits on their characteristic sizes in the order of 20-30\rg.

The objects explored here span a large range in mass and Eddington
fraction; however, in all cases our selection criteria ($L\gtrsim 10^{-2}L_{Edd}$) was such that  the systems are thought to be in
states where the accretion disk extends down to the ISCO \citep[][and references therein]{reis20121650};. Based on the
 presence of highly compact coronae, we postulate that the
Seyferts and quasars presented here are in spectral states analogous to the
hard/intermediate and the soft/intermediate states of stellar mass BHBs.

The compactness of the coronae in these sources strongly constrains
models of Comptonisation, with magnetic reconnection/flares
\citep[e.g.][]{MerloniFabian2001} or failed jet
models \citep[e.g.][]{Ghisellini04jets} strongly favoured over models
which predict larger, extended coronae
\citep[e.g.][]{HaardtMaraschi91}.  The small distance between the
corona and the accretion disk further suggests that models where the
energy is dissipated in the innermost regions or possibly even within
the plunging regions, are favoured. Such a scenario was presented by
\citet{Hirose04}, where the corona is described as a region of smooth
magnetic field lines increasing in strength toward small radii and
being strongest for maximally rotating BHs.  More recently,
global MHD simulations were used to show that
the vast majority of the coronal emission comes from a relatively
small volume of space \citep{Schnittman2012states}, consistent with observational results presented in  this work.

We also note  that the base of the jet in M87 has recently been
spatially resolved to be $\lesssim11$\rg\ in size
\citep{Doeleman2012Sci}.  If this is  associated with the coronae
-- as is the case in some models \citep[e.g.][]{FalckeBiermann1999} -- then this too argues for a highly compact
corona.

A number of  arguments have so far eluded discussion and shall
be mentioned here only briefly. First, the spectral fits to broad,
relativistic lines profiles often require emissivities that are
indicative of a very compact corona. This has been seen in both
stellar-mass as well as SMBH and is often interpreted as the
gravitational bending of light toward the inner regions of the disk
\citep[][]{Miniu04}.  We have chosen not to use this line of evidence as
 argument for a compact corona as it has recently been shown
that these results could potentially be dependent on the prior
assumption that the corona is inherently compact \citep{Dauser2013}.

The presence of compact coronae are also often inferred by the
existence of high frequency QPOs whose frequencies are similar to that
expected from the Keplerian frequencies in the inner few gravitational radii from
the black hole. QPOs are known to be stronger at higher energies which
has prompted the association of the corona as the main instrument in
modulating the signals. Indeed, as mentioned above,  QPOs have been
seen in RE J1034+396 and in the tidal disruption source Swift~J164449.3+573451 \citep{Reis2012qpo}, and in both cases the QPOs  have been associated with the coronae  as the emission from the disk peaks in the optical/UV. However, the
drawback in inferring the properties of the corona from the QPO, is that to date there is no self-consistent model that can fully explain
the observed behaviour in a wide variety of sources.

In summary, current evidence -- based on both imaging and timing
techniques -- strongly favours a compact corona located a few
gravitational radii above the accretion disk.  Whether the corona follows a slab, spherical, conical or patchy geometry, or whether it
contains a purely thermal, non-thermal or hybrid population of
electrons is still unknown.  However, future emission models should
ensure that the majority of the energy dissipation occurs in a compact
region close to the central black hole, if they are to be physical. Of course,  both X-ray macrolensing and soft-lag techniques are still in their early days, and  the application of these techniques in determining the size and location of  coronal regions should be checked against a much larger sample in the near future.

\section{Acknowledgements}
RCR is supported by NASA through the Einstein Fellowship Program, grant number PF1-120087. RCR thank Monica Valluri for the updated mass for NGC~ 4151 as well as Ed. Cackett and Dom Walton for helpful comments.

\begin{thebibliography}{52}
\expandafter\ifx\csname natexlab\endcsname\relax\def\natexlab#1{#1}\fi

\bibitem[{Bentz} et~al.(2006){Bentz}, {Denney}, {Cackett}
  et~al.]{Bentzngc41512006}
{Bentz} M.~C., {Denney} K.~D., {Cackett} E.~M., et~al., 2006, \apj, 651, 775

\bibitem[{Berti} \& {Volonteri}(2008)]{BertiVolonteri2008}
{Berti} E., {Volonteri} M., 2008, \apj, 684, 822

\bibitem[{Blackburne} et~al.(2006){Blackburne}, {Pooley} \&
  {Rappaport}]{BlackburnePooley2006}
{Blackburne} J.~A., {Pooley} D., {Rappaport} S., 2006, \apj, 640, 569

\bibitem[{Cackett} et~al.(2012){Cackett}, {Fabian}, {Zoghbi}, {Kara},
  {Reynolds} \& {Uttley}]{Cackett2012ESOlag}
{Cackett} E.~M., {Fabian} A.~C., {Zoghbi} A., {Kara} E., {Reynolds} C.,
  {Uttley} P., 2012, ArXiv e-prints

\bibitem[{Chartas} et~al.(2009){Chartas}, {Kochanek}, {Dai}, {Poindexter} \&
  {Garmire}]{size1104}
{Chartas} G., {Kochanek} C.~S., {Dai} X., {Poindexter} S., {Garmire} G., 2009,
  \apj, 693, 174

\bibitem[{Dalal} \& {Kochanek}(2002)]{DalalKochanek2002}
{Dalal} N., {Kochanek} C.~S., 2002, \apj, 572, 25

\bibitem[{Dauser} et~al.(2013){Dauser}, {Garcia}, {Wilms} et~al.]{Dauser2013}
{Dauser} T., {Garcia} J., {Wilms} J., et~al., 2013, ArXiv e-prints

\bibitem[{De Marco} et~al.(2012){De Marco}, {Ponti}, {Cappi}
  et~al.]{DeMarcolags}
{De Marco} B., {Ponti} G., {Cappi} M., et~al., 2012, ArXiv e-prints


\bibitem[{Doeleman} et~al.(2012){Doeleman}, {Fish}, {Schenck}
  et~al.]{Doeleman2012Sci}
{Doeleman} S.~S., {Fish} V.~L., {Schenck} D.~E., et~al., 2012, Science, 338,
  355

\bibitem[{Elvis} et~al.(1978){Elvis}, {Maccacaro}, {Wilson} et~al.]{Elvis1978}
{Elvis} M., {Maccacaro} T., {Wilson} A.~S., et~al., 1978, \mnras, 183, 129

\bibitem[{Fabian} et~al.(2012){Fabian}, {Kara}, {Walton}
  et~al.]{Fabian2012iras13224}
{Fabian} A.~C., {Kara} E., {Walton} D.~J., et~al., 2012, ArXiv e-prints

\bibitem[{Fabian} et~al.(2009){Fabian}, {Zoghbi}, {Ross} et~al.]{FabZog09}
{Fabian} A.~C., {Zoghbi} A., {Ross} R.~R., et~al., 2009, \nat, 459, 540

\bibitem[{Falcke} \& {Biermann}(1999)]{FalckeBiermann1999}
{Falcke} H., {Biermann} P.~L., 1999, \aap, 342, 49

\bibitem[{Ghisellini} et~al.(2004){Ghisellini}, {Haardt} \&
  {Matt}]{Ghisellini04jets}
{Ghisellini} G., {Haardt} F., {Matt} G., 2004, \aap, 413, 535

\bibitem[{Gierli{\'n}ski} et~al.(2008){Gierli{\'n}ski}, {Middleton}, {Ward} \&
  {Done}]{rej1034qpo}
{Gierli{\'n}ski} M., {Middleton} M., {Ward} M., {Done} C., 2008, \nat, 455, 369

\bibitem[{Haardt} \& {Maraschi}(1991)]{HaardtMaraschi91}
{Haardt} F., {Maraschi} L., 1991, \apjl, 380, L51

\bibitem[{Hirose} et~al.(2004){Hirose}, {Krolik}, {De Villiers} \&
  {Hawley}]{Hirose04}
{Hirose} S., {Krolik} J.~H., {De Villiers} J.-P., {Hawley} J.~F., 2004, \apj,
  606, 1083

\bibitem[{Irwin} et~al.(1989){Irwin}, {Webster}, {Hewett}, {Corrigan} \&
  {Jedrzejewski}]{Irwin1989}
{Irwin} M.~J., {Webster} R.~L., {Hewett} P.~C., {Corrigan} R.~T.,
  {Jedrzejewski} R.~I., 1989, \aj, 98, 1989

\bibitem[{Johannsen} \& {Psaltis}(2011)]{Johannsen2011nohair}
{Johannsen} T., {Psaltis} D., 2011, \prd, 83, 12, 124015

\bibitem[{Kara} et~al.(2013){Kara}, {Fabian}, {Cackett}, {Miniutti} \&
  {Uttley}]{Kara2013iras}
{Kara} E., {Fabian} A.~C., {Cackett} E.~M., {Miniutti} G., {Uttley} P., 2013,
  \mnras, 430, 1408

\bibitem[{McHardy} et~al.(2006){McHardy}, {Koerding}, {Knigge}, {Uttley} \&
  {Fender}]{McHardy2006}
{McHardy} I.~M., {Koerding} E., {Knigge} C., {Uttley} P., {Fender} R.~P., 2006,
  \nat, 444, 730

\bibitem[{Merloni} \& {Fabian}(2001)]{MerloniFabian2001}
{Merloni} A., {Fabian} A.~C., 2001, \mnras, 321, 549

\bibitem[{Merloni} et~al.(2003){Merloni}, {Heinz} \& {di
  Matteo}]{fundamentalplane}
{Merloni} A., {Heinz} S., {di Matteo} T., 2003, \mnras, 345, 1057

\bibitem[{Metcalf} \& {Zhao}(2002)]{MetcalZhaof2002}
{Metcalf} R.~B., {Zhao} H., 2002, \apjl, 567, L5

\bibitem[{Miller}(2007)]{miller07review}
{Miller} J.~M., 2007, \araa, 45, 441

\bibitem[{Miller} et~al.(2011){Miller}, {Miller} \& {Reynolds}]{millersn11}
{Miller} J.~M., {Miller} M.~C., {Reynolds} C.~S., 2011, \apjl, 731, L5+

\bibitem[{Miller} et~al.(2010){Miller}, {Turner}, {Reeves} \&
  {Braito}]{MillerTurner2010lag}
{Miller} L., {Turner} T.~J., {Reeves} J.~N., {Braito} V., 2010, \mnras, 408,
  1928

\bibitem[{Miniutti} \& {Fabian}(2004)]{Miniu04}
{Miniutti} G., {Fabian} A.~C., 2004, \mnras, 349, 1435

\bibitem[{Morgan} et~al.(2008){Morgan}, {Kochanek}, {Dai}, {Morgan} \&
  {Falco}]{sizepg1115}
{Morgan} C.~W., {Kochanek} C.~S., {Dai} X., {Morgan} N.~D., {Falco} E.~E.,
  2008, \apj, 689, 755

\bibitem[{Morgan} et~al.(2010){Morgan}, {Kochanek}, {Morgan} \&
  {Falco}]{MorganKochanek2010}
{Morgan} C.~W., {Kochanek} C.~S., {Morgan} N.~D., {Falco} E.~E., 2010, \apj,
  712, 1129

\bibitem[{Mortonson} et~al.(2005){Mortonson}, {Schechter} \&
  {Wambsganss}]{Mortonson2005}
{Mortonson} M.~J., {Schechter} P.~L., {Wambsganss} J., 2005, \apj, 628, 594

\bibitem[{Mosquera} et~al.(2013){Mosquera}, {Kochanek}, {Chen}, {Dai},
  {Blackburne} \& {Chartas}]{Mosquera2013}
{Mosquera} A.~M., {Kochanek} C.~S., {Chen} B., {Dai} X., {Blackburne} J.~A.,
  {Chartas} G., 2013, ArXiv e-prints

\bibitem[{Narayan} \& {McClintock}(2012)]{NarayanMcClintock2012}
{Narayan} R., {McClintock} J.~E., 2012, \mnras, 419, L69

\bibitem[{Narayan} \& {Yi}(1994)]{NarayanYi1994}
{Narayan} R., {Yi} I., 1994, \apjl, 428, L13

\bibitem[{Onken} et~al.(2007){Onken}, {Valluri}, {Peterson}
  et~al.]{Onkenngc41512007}
{Onken} C.~A., {Valluri} M., {Peterson} B.~M., et~al., 2007, \apj, 670, 105

\bibitem[{Peng} et~al.(2006){Peng}, {Impey}, {Rix} et~al.]{Peng2006amass}
{Peng} C.~Y., {Impey} C.~D., {Rix} H.-W., et~al., 2006, \apj, 649, 616

\bibitem[{Pooley} et~al.(2007){Pooley}, {Blackburne}, {Rappaport} \&
  {Schechter}]{PooleyBlackburne2007}
{Pooley} D., {Blackburne} J.~A., {Rappaport} S., {Schechter} P.~L., 2007, \apj,
  661, 19

\bibitem[{Reis} et~al.(2012){Reis}, {Miller}, {Reynolds} et~al.]{Reis2012qpo}
{Reis} R.~C., {Miller} J.~M., {Reynolds} M.~T., et~al., 2012, Science, 337, 949

\bibitem[{Reis} et~al.(2013){Reis}, {Miller}, {Reynolds} et~al.]{reis20121650}
{Reis} R.~C., {Miller} J.~M., {Reynolds} M.~T., et~al., 2013, \apj, 763, 48

\bibitem[{Schnittman} et~al.(2012){Schnittman}, {Krolik} \&
  {Noble}]{Schnittman2012states}
{Schnittman} J.~D., {Krolik} J.~H., {Noble} S.~C., 2012, ArXiv e-prints

\bibitem[{Sluse} et~al.(2012){Sluse}, {Hutsem{\'e}kers}, {Courbin}, {Meylan} \&
  {Wambsganss}]{Sluse2012mass}
{Sluse} D., {Hutsem{\'e}kers} D., {Courbin} F., {Meylan} G., {Wambsganss} J.,
  2012, \aap, 544, A62

\bibitem[{Valluri} et~al.(2004){Valluri}, {Merritt} \&
  {Emsellem}]{vallurietal04}
{Valluri} M., {Merritt} D., {Emsellem} E., 2004, \apj, 602, 66

\bibitem[{Walton} et~al.(2013){Walton}, {Nardini}, {Fabian}, {Gallo} \&
  {Reis}]{Waltonreisspin2013}
{Walton} D.~J., {Nardini} E., {Fabian} A.~C., {Gallo} L.~C., {Reis} R.~C.,
  2013, \mnras, 428, 2901

\bibitem[{Wambsganss}(2006)]{Wambsganss2006}
{Wambsganss} J., 2006, in { Saas-Fee Advanced Course 33: Gravitational Lensing:
  Strong, Weak and Micro\/}, edited by G.~{Meylan}, P.~{Jetzer}, P.~{North},
  P.~{Schneider}, C.~S. {Kochanek}, J.~{Wambsganss},  453--540

\bibitem[{White} \& {Holt}(1982)]{White1982coronae}
{White} N.~E., {Holt} S.~S., 1982, \apj, 257, 318

\bibitem[{Wilkins} \& {Fabian}(2013)]{WilkinsFabian2013}
{Wilkins} D.~R., {Fabian} A.~C., 2013, \mnras,  602

\bibitem[{Zdziarski} et~al.(1999){Zdziarski}, {Lubi{\'n}ski} \&
  {Smith}]{Zdziarski1999}
{Zdziarski} A.~A., {Lubi{\'n}ski} P., {Smith} D.~A., 1999, \mnras, 303, L11

\bibitem[{Zhou} \& {Wang}(2005)]{Zhoumasses2005}
{Zhou} X.-L., {Wang} J.-M., 2005, \apjl, 618, L83

\bibitem[{Zhou} et~al.(2010){Zhou}, {Zhang}, {Wang} \&
  {Zhu}]{zhouzhangmasses2010}
{Zhou} X.-L., {Zhang} S.-N., {Wang} D.-X., {Zhu} L., 2010, \apj, 710, 16

\bibitem[{Zoghbi} et~al.(2012){Zoghbi}, {Fabian}, {Reynolds} \&
  {Cackett}]{Zoghbi2012lag}
{Zoghbi} A., {Fabian} A.~C., {Reynolds} C.~S., {Cackett} E.~M., 2012, \mnras,
  422, 129

\end{thebibliography}
%

\end{document}